# ON THE PROBLEM OF THE MAGNETIC MASS

O. Philipsen

Theory Group, DESY
Notkestr. 85
D-22603 Hamburg, Germany

## 1. INTRODUCTION

A complete perturbative description of non-abelian field theories at finite temperature is prohibited due to the occurrence of infrared divergencies for massless fields and the corresponding breakdown of perturbation theory. In particular, this renders impossible a discussion of the electroweak phase transition in the immediate vicinity of the critical temperature, as well as an evaluation of the effective potential in the symmetric phase. A potential cure for this problem is the dynamical generation of a "magnetic mass", which has to be of the form $\sim g^2 T$, but whose coefficient cannot be calculated perturbatively, since it receives contributions from all orders of perturbation theory.[1]

Recently an estimate of this coefficient was attempted[2] by studying a set of gap equations for the full propagators of the theory. In this approach one rearranges the perturbation series by adding and subtracting mass counterterms to the action,

$$S = S_0 + \delta S_m + S_{int} - \delta S_m , \qquad (1)$$

and expanding in the modified interaction term $S_{int} - \delta S_m$ with propagators derived from $S_0 + \delta S_m$. If the up to now arbitrary counterterms $\delta S_m$ are chosen to represent the radiative corrections to the masses of the corresponding fields, one obtains propagators containing full masses $m^2 = m_0^2 + \delta m^2$, which are to be used in the evaluation of loop diagrams in the modified perturbation theory. The counterterm $-\delta S_m$ in the new interaction term ensures that the higher order contributions, which have been resummed into the propagator, do not contribute twice when calculations in higher orders are done. The value of the mass corrections $\delta m^2$ can be determined implicitly and self-consistently by calculating the self-energies of the fields in the modified perturbative scheme using full propagators. This corresponds to an infinite iteration of one-loop mass corrections and leads to a gap equation for the full masses $m$.[3] In the case of the SU(2)-Higgs model the authors of Refs. 2 considered the gap equation

$$m_T^2 = \Pi_T(p_0 = 0, p \to 0) \qquad (2)$$

mass $m_{T0} = 0$), and they found in Landau gauge

$$m_T^2 = \frac{m_T}{3\pi} g^2 T , \qquad (3)$$

yielding a nonvanishing magnetic mass $m_T = g^2 T/(3\pi)$.

Most equilibrium properties of field theories at very high temperature can be investigated by studying the corresponding dimensionally reduced field theory.[4] It is therefore instructive to examine, if the dynamical generation of a mass for the spatial components of the vector fields can also be understood in the three-dimensional theory.[5]

## 2. PURE GAUGE THEORY IN 3 DIMENSIONS

Consider a pure gauge theory in three euclidean dimensions, where the gauge fields are known to remain massless at every finite order perturbation theory. In order to study non-perturbative dynamical mass generation a mass counterterm $A_\mu^a A_\mu^a m^2/2$ is added and subtracted to the action following the rationale explained in the last section. After covariant gauge fixing and including the ghost term the Lagrangian density relevant for studying the one loop self-energy of the gauge fields is given by

$$\mathcal{L} = \tfrac{1}{4} F_{\mu\nu}^a F_{\mu\nu}^a + \tfrac{m^2}{2} A_\mu^a A_\mu^a + \tfrac{1}{2\xi}(\partial_\mu A_\mu^a)^2 - \bar{c}_a \partial_\mu \left[ \delta_{ab} \partial_\mu + g\varepsilon_{abc} A_\mu^c \right] c_b . \qquad (4)$$

In three dimensions there are only two linearly independent tensors with two Lorentz indices which can be chosen to be the transverse and longitudinal projectors,

$$P_{L\mu\nu} = \frac{p_\mu p_\nu}{p^2} ,$$
$$P_{T\mu\nu} = \delta_{\mu\nu} - \frac{p_\mu p_\nu}{p^2} . \qquad (5)$$

The propagator and the self-energy can thus be decomposed,

$$D_{\mu\nu}^{ab}(p) = \delta_{ab}[D_T(p)P_{T\mu\nu} + D_L(p)P_{L\mu\nu}] ,$$
$$\Pi_{\mu\nu}^{ab}(p) = \delta_{ab}[\Pi_T(p)P_{T\mu\nu} + \Pi_L(p)P_{L\mu\nu}] , \qquad (6)$$

yielding the full propagator

$$\tilde{D}_{\mu\nu}^{ab}(p) = \sum_{n=0}^{\infty} D(p)[\Pi(p)D(p)]^n$$
$$= \delta_{ab} \left[ \frac{P_{T\mu\nu}}{p^2 - \Pi_T(p)} + \frac{P_{L\mu\nu}}{p^2 - \Pi_L(p)} \right] . \qquad (7)$$

In order to find the magnetic mass the transverse self-energy $\Pi_T$ has to be calculated with the propagators following from (4). If the magnetic mass determined in this way is to have a physical meaning instead of being just a technical quantity appearing in calculations, it has to be independent of the gauge fixing parameter. We therefore do the calculation in a general covariant gauge in order to check the gauge dependence explicitly. Thus, we study the gap equation

$$m^2 = -\Pi_T(p \to 0, m, \xi) , \qquad (8)$$

which is shown diagrammatically in Fig. 1. Removing linearly divergent terms by means of dimensional regularization we obtain

$$m^2 = -g^2 \tfrac{m}{\pi} \left[ -\tfrac{5}{6} + \tfrac{1}{3}\xi^{3/2} + \tfrac{3-5\xi+2\xi^{5/2}}{6(1-\xi)} \right] . \qquad (9)$$

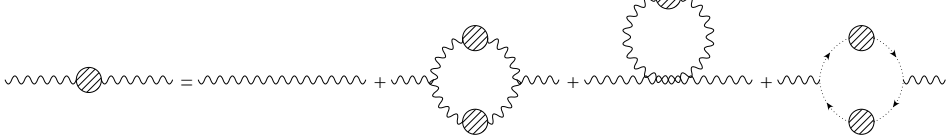

**Figure 1.** Gap equation for the self-consistent determination of the full mass correction for the propagator. The blobs on the lines indicate full propagators.

In Landau gauge, $\xi = 0$, and with the replacement $g^2 \to g^2 T$ this reduces to equation (3). For four-dimensional theories at $T \neq 0$ we can therefore conclude that the generation of a magnetic mass as described by gap equations at one-loop order is a feature of the Matsubara zero modes.

However, the mass term generated in this way is gauge dependent. In a study of the Debye mass it has been pointed out by Rebhan[6] that, in order to obtain gauge invariant results, one has to define the mass by the position of the pole in the propagator rather than by the zero momentum limit of the self-energy. Adopting this prescription one has to consider the gap equation

$$m^2 = -\Pi_T(p^2 = -m^2, m, \xi) \, , \tag{10}$$

instead of (8). However, the corresponding expression for the transverse self-energy still depends on $\xi$.

## 3. MASSIVE YANG-MILLS FIELDS IN 3 DIMENSIONS

A gauge invariant way to implement mass terms for non-abelian gauge fields is provided by the mechanism of spontaneous symmetry breaking. Consider the gauge theory supplemented by a scalar field,

$$\mathcal{L} = \tfrac{1}{4} F^a_{\mu\nu} F^a_{\mu\nu} + \mathrm{Tr}(D_\mu \Phi)^\dagger (D_\mu \Phi) + \mu^2 \mathrm{Tr}(\phi^\dagger \phi) + \tfrac{\lambda}{6}\left[\mathrm{Tr}(\phi^\dagger \phi)^2\right] \, , \tag{11}$$

where $\Phi$ is a complex $2 \times 2$ matrix field transforming as the (1/2,1/2) representation of SU(2)×SU(2). This field develops a nonvanishing vacuum expectation value $\langle \Phi \rangle = v$, and can be parametrized by four real fields,

$$\Phi = \tfrac{1}{2}(\sigma + i\tau^a \pi^a) \, . \tag{12}$$

Inserting this into (11) leads to the $\sigma$-model. Upon imposing the condition $\sigma^2 + (\pi^a)^2 = v^2$ and sending $\lambda, \mu \to \infty$ with $v$ fixed the $\sigma$ field decouples and one obtains the non-linear $\sigma$-model,

$$\mathcal{L} = \tfrac{1}{4} F^a_{\mu\nu} F^a_{\mu\nu} + \tfrac{g^2 v^2}{8} A^a_\mu A^a_\mu + \tfrac{1}{2}(\partial_\mu \pi^a)^2 + \tfrac{gv}{2}(\partial_\mu \pi^a) A^a_\mu - \tfrac{g}{2}\varepsilon_{abc}\partial_\mu \pi^a A^b_\mu \pi^c$$
$$+ O(\pi^4, \pi^3 A, ...) \, . \tag{13}$$

where the gauge field has acquired a mass $m_0 = gv/2$. We now perform the same procedure for this model as in the last section, the difference being that in this case we start out with a tree level mass $m_0 \neq 0$, which has to be sent to zero at the end of the calculation. To this Lagrangian we add the mass correction for the gauge fields, the $R_\xi$ gauge fixing term

$$\mathcal{L}_{gf} = \tfrac{1}{2\xi}(\partial_\mu A^a_\mu + m\xi \pi^a)^2 \, ,$$

and the corresponding ghost term to arrive at

$$\mathcal{L} = \tfrac{1}{4} F^a_{\mu\nu} F^a_{\mu\nu} + \tfrac{m^2}{2} A^a_\mu A^a_\mu + \tfrac{1}{2\xi}(\partial_\mu A^a_\mu)^2 + \tfrac{1}{2}(\partial_\mu \pi^a)^2 + \tfrac{1}{2} m^2 \xi \pi^a \pi^a - \tfrac{g}{2} \varepsilon_{abc} \partial_\mu \pi^a A^b_\mu \pi^c$$
$$- \bar{c}_a \left[ \partial_\mu \left( \delta_{ab} \partial_\mu + g \varepsilon_{abc} A^c_\mu \right) + \tfrac{g^2 v}{4} \xi \varepsilon_{abc} \pi^c - m^2 \xi \delta_{ab} \right] c_b \; . \tag{14}$$

The Goldstone fields $\pi^a$ are unphysical and can be gauged away by going to the unitary gauge, in which (14) reduces to a theory describing a massive vector field,

$$\mathcal{L}_{unitary} = \tfrac{1}{4} F^a_{\mu\nu} F^a_{\mu\nu} + \tfrac{m^2}{2} A^a_\mu A^a_\mu \; . \tag{15}$$

The self-energy of the gauge fields now gets an additional contribution from the Goldstone loop, Fig. 2.

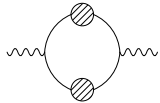

**Figure 2.** Contribution of the Goldstone modes to the gap equation for the model (14).

We have derived the gap equations (8) and (10) for this model. In the first case we recover equation (9) with an additional term $\sim \sqrt{\xi}$ from the Goldstone and ghost diagrams, leaving the expression gauge dependent. However, for the gap equation (10) we find in the limit $m_0 = 0$

$$m^2 = g^2 \tfrac{m}{\pi} \left[ \tfrac{63}{64} \ln 3 - \tfrac{3}{16} \right] \; . \tag{16}$$

This gap equation is manifestly gauge invariant and has, besides $m = 0$, the non-trivial solution

$$m = 0.28 g^2 \; . \tag{17}$$

In conclusion it was found that dynamical generation of a transverse gauge field mass as it can be investigated by the study of gap equations occurs in purely three-dimensional theories and hence, in the case of four-dimensional theories at $T \neq 0$, can be attributed to the static modes. We found a gauge invariant pole of the resummed gauge field propagator of the non-linear $\sigma$-model in the limit of vanishing tree level mass. If one interprets this model as the high temperature limit of four-dimensional SU(2) pure gauge theory, the value (17) for the magnetic mass is consistent with the results of other non-perturbative studies.[7] In order to understand the precise meaning of the above results for physical, four-dimensional theories further studies are required. An investigation of the connection to the linear $\sigma$-model as well as a more systematic discussion of renormalization in the approach presented here is currently in progress.[5]